\documentstyle[emulateapj,epsfig]{article}

\def\ifm#1{\relax\ifmmode#1\else$\mathsurround=0pt #1$\fi}
\def\kms{\ifmmode\,{\rm km}\,{\rm s}^{-1}\else km$\,$s$^{-1}$\fi}

\def\hkpc{\,{\rm h^{-1}kpc}}

\def\msun{M_{\odot}}
\def\hmsun{h^{-1}\msun}
\def\ltsima{$\; \buildrel < \over \sim \;$}
\def\lsim{\lower.5ex\hbox{\ltsima}}
\def\gtsima{$\; \buildrel > \over \sim \;$}
\def\gsim{\lower.5ex\hbox{\gtsima}}

\def\lcdm{$\Lambda$CDM}
\def\om{\Omega_{\rm m}}
\def\olam{\Omega_{\Lambda}}

\newenvironment{inlinefigure}{
\def\@captype{figure}
\noindent\begin{minipage}{0.999\linewidth}\begin{center}}
{\end{center}\end{minipage}\smallskip}

\begin{document}
\slugcomment{{\em submitted to Astrophysical Journal Letters}}

\lefthead{CAN PHOTOIONIZATION SQUELCHING RESOLVE THE SUB-STRUCTURE CRISIS?}
\righthead{SOMERVILLE} 
\title{Can Photoionization Squelching Resolve the Substructure Crisis?}

%-------------------------------------------------------------
\author{Rachel S. Somerville \altaffilmark{1}}
\affil {Institute of Astronomy, University of Cambridge, 
Madingley Rd., Cambridge, CB3 0HA, UK}

\altaffiltext{1}{rachel@ast.cam.ac.uk}

\begin{abstract}
Cold Dark Matter theory predicts that the Local Group should contain
many more dwarf-sized objects than the observed number of dwarf
galaxies --- the so-called sub-structure problem. We investigate
whether the suppression of star formation in these small objects due
to the presence of a photoionizing background can resolve the problem.
We make use of results from recent hydrodynamic simulations to build a
recipe for the suppression of gas infall into semi-analytic galaxy
formation models, and use these to predict the luminosity function of
dwarf galaxies in the Local Group. In the models without
photoionization ``squelching'', we predict a large excess of faint
dwarf galaxies compared with the observed number in the Local Group
--- thus, the usual recipe for supernovae feedback used in
semi-analytic models does not solve the sub-structure problem on its
own. When we include photoionization squelching, we find good
agreement with the observations. We have neglected tidal destruction,
which probably further reduces the number of dwarf galaxies.  We
conclude that photoionizing squelching easily solves the sub-structure
problem. In fact, it is likely that once this effect is taken into
account, models with reduced small-scale power (e.g. Warm Dark Matter)
would \emph{underproduce} dwarf galaxies.
\end{abstract}

\subjectheadings{cosmology: dark matter --- galaxies:
formation --- galaxies: dwarf}

%=======================
% 1
\section{Introduction}
\label{sec:intro}
%=======================
Cold Dark Matter (CDM) models with scale-invariant initial fluctuation
spectra seem to provide a remarkably successful framework for
explaining a broad range of observations pertaining to structure on
scales larger than $\simeq 1$ Mpc.  Only recently have N-body
techniques achieved the dynamic range necessary to resolve
sub-structures within other bound, virialized structures. This new
generation of very high resolution simulations has demonstrated that
the presence of numerous sub-structures is an unavoidable consequence
within the cosmological constant ($\Lambda$) dominated ``\lcdm''
models that are now very widely used. Such simulations predict that
dark matter halos with the mass of the halos of the Milky Way or M31
contain within their virial radius several hundred objects with
internal velocities similar to those of observed Local Group dwarf
galaxies (Klypin et al. 1999 (KKVP); Moore et al. 1999 (M99)), whereas
only about 40 such galaxies are known in the Local Group.  This result
has come to be known as the ``sub-structure problem'', and is a direct
consequence of the relatively large amount of power on small scales in
``standard'' \lcdm\ models.

A variety of solutions have been proposed which solve the problem by
modifying some basic tenant of the usual CDM canon. These include
reducing the small-scale power by appealing to inflationary models
with broken scale invariance (Kamionkowski \& Liddle 2000), or
changing the nature of the dark matter, e.g. Warm Dark Matter (Hogan
\& Dalcanton 2000; Bode et al. 2001), 
Self-interacting Dark Matter (Spergel \& Steinhardt 2000), or Strongly
Annihilating Dark Matter (Kaplinghat, Knox \& Turner 2000). It
appears, however, that each of these solutions may violate other
constraints (Barkana, Haiman, \& Ostriker 2001; Gnedin \& Ostriker
2001; Hui 2001; Miralda-Escud\'{e} 2000).  Therefore, in this paper,
we investigate a less exotic solution.

It has long been appreciated that the accretion and cooling of gas by
low mass halos will be ``squelched'' in the presence of a strong
photoionizing background. This process has been studied by many
authors using analytic approaches (Ikeuchi 1986; Rees 1986; Babul \&
Rees 1992; Efstathiou 1992; Shapiro, Giroux, \& Babul 1994), 1-D
hydrodynamic simulations (Haiman, Thoul, \& Loeb 1996; Thoul \&
Weinberg 1996) and full 3-D hydro simulations (Quinn, Katz, \&
Efstathiou 1996; Weinberg, Hernquist, \& Katz 1997; Navarro \&
Steinmetz 1997; Gnedin 2000). These studies concur in the conclusion
that after reionization, gas accretion is suppressed in halos with
virial temperature less than some characteristic value. The Jeans Mass
is an obvious candidate for this critical mass scale (e.g. Rees
1986). However, because dark matter halos in CDM models have complex
and varied formation histories, it is clear that some sort of
time-averaged Jeans Mass is needed to determine the gas content of any
given halo at any given time. The recent analysis of Gnedin (2000;
G00), based on hydrodynamic simulations, suggests that the `filtering
mass', which corresponds to the length scale over which baryonic
perturbations are smoothed in linear theory, provides a good
description of this time-averaged, effective Jeans Mass. The net
result is that halos with virial velocities $\lsim$ 30--50 km/s are
able to collect a substantial amount of gas only if they collapse
before the universe is reionized.

Bullock, Kravtsov, \& Weinberg (2000; BKW00) investigated whether this
effect could resolve the sub-structure problem, using a model based on
estimates of halo collapse redshifts obtained from an extension of the
Press-Schechter (1974) approximation, combined with constraints from
N-body results. They concluded that a combination of tidal destruction
and photoionization squelching could very naturally solve the
sub-structure problem in the local group. However, their model did not
treat gas physics or star formation explicitly, instead relying on a
few phenomenological parameters to determine which halos would host
observable galaxies. Chiu, Gnedin \& Ostriker (2001) also concluded
that photoionization squelching would reduce the slope of the galactic
mass-function on the scale of dwarf galaxy halos, using scaling
arguments based on the Press-Schechter approximation and results from
the simulations of G00. It remains worthwhile, however, to investigate
this process within the more detailed framework of semi-analytic
galaxy formation modeling (e.g. Kauffmann, White, \& Guiderdoni 1993;
Somerville \& Primack 1999; Cole et al. 2000).

In this paper, we use the recipe provided by Gnedin (2000), based on
hydrodynamic simulations, to incorporate the effects of
photoionization squelching into a semi-analytic model of galaxy
formation. We examine the effect on the luminosity function of dwarf
galaxies in Milky Way-sized halos and re-assess whether this process
can solve the sub-structure problem.  Our models are based on Monte
Carlo merger trees and include treatments of sub-halo merging, gas
cooling, star formation, supernovae feedback, chemical enrichment, and
stellar population synthesis. In all the models presented here, we
adopt a ``standard'' \lcdm\ cosmology with $\om=0.3$, $\olam=0.7$, $H_0
= 70$ km/s/Mpc, and $\sigma_8=1$.

\section{Semi-analytic Models}
\label{sec:models}
\subsection{Merger Trees and Sub-structure}
%redo plot in BW?
Our models are based on Monte Carlo realizations of merger trees
constructed using the extended Press-Schechter formalism and the
method of Somerville \& Kolatt (1999). We track the decay of the
orbits of sub-halos due to dynamical friction after being subsumed
within a larger halo, and when a sub-halo reaches the center of the
host halo, it is assumed to merge with the central object (for details
see Somerville \& Primack 1999; hereafter SP). We neglect tidal
destruction of sub-halos, and also satellite-satellite merging. We
would first like to check that this approximate treatment of the
survival of substructure is in agreement with N-body
simulations. Fig.~\ref{fig:submf} shows the cumulative velocity
function predicted by the semi-analytic models for sub-halos that
reside within the virial radius of a host halo with a mass of $10^{12}
\msun$. This is compared with the velocity function of sub-halos within 200
$\hkpc$, for halos of similar mass from the N-body simulations of M99
and from recent simulations produced using a new version of the
Adaptive Refinement Tree (ART) code (Kravtsov, Klypin, \& Khoklov
1997) with variable mass particles.  These simulations, which we will
refer to here as `the ART simulations', are similar to those presented
in KKVP99, but have higher mass resolution. The results are binned in
terms of $V_{\rm max}$, the circular velocity at the peak of the
rotation curve, which is the quantity that is most robust for
sub-halos identified in the simulations.  We use the results of
Bullock et al. (2001) to calculate $V_{\rm max}$ for the halos in the
semi-analytic merger trees as a function of their mass and collapse
redshift. The semi-analytic trees have been run with the same mass
resolution as the N-body simulations (the smallest halos have a mass
of $10^{7} \hmsun$).

\begin{inlinefigure}
\begin{center}
\resizebox{\textwidth}{!}{\includegraphics{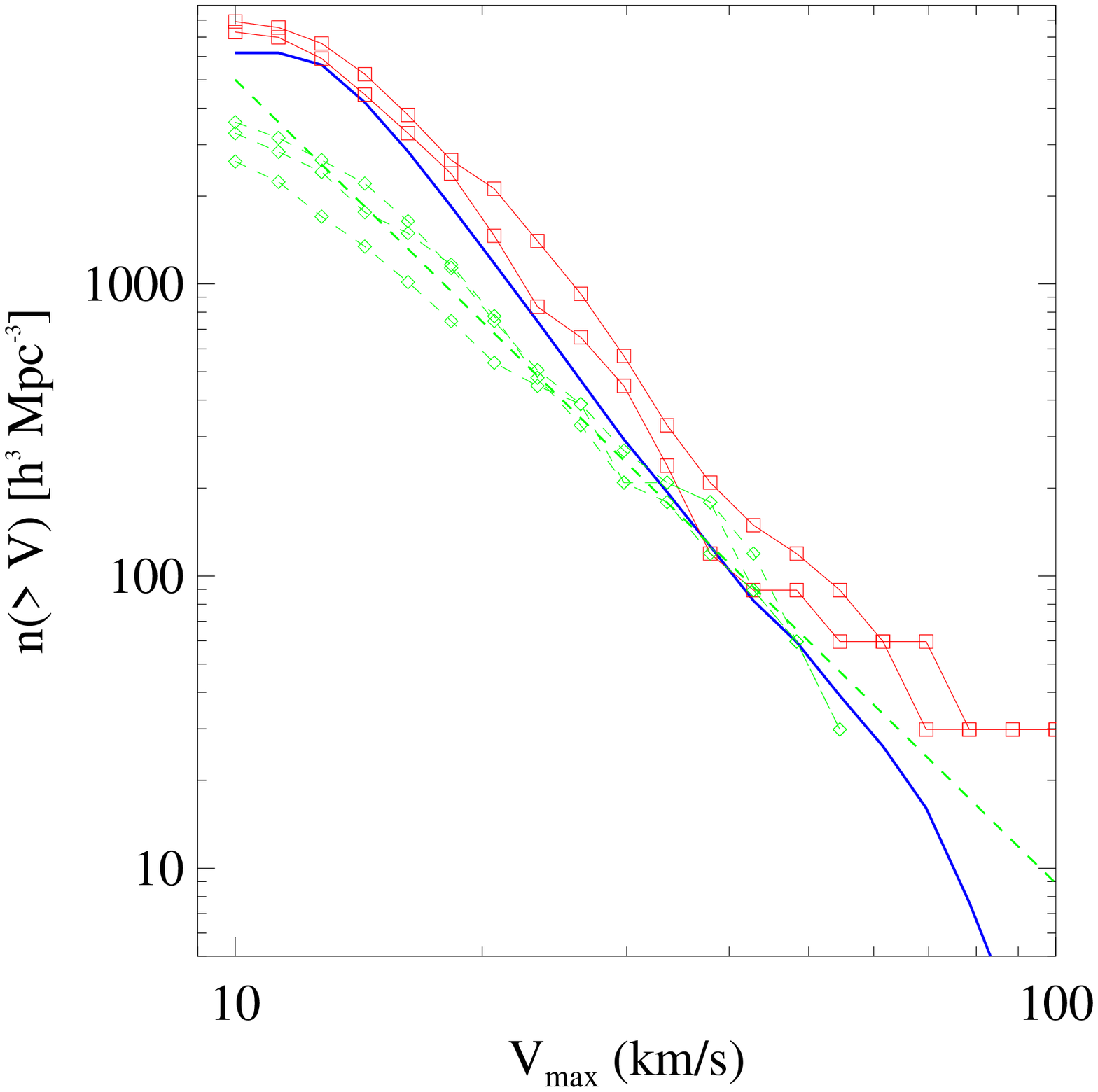}}
\end{center}
\figcaption{The volume density of sub-halos, as a function of their
maximum circular velocity $V_{\rm max}$. The thick solid line shows
the results from many realizations of a semi-analytic merger tree
(normalized to the volume within the virial radius). Thin solid lines
with squares show the results for two halos from the simulations of
M99. Dashed lines with diamonds show the results from the ART
simulations (see text), and the straight dashed line shows the
power-law fit to the simulations presented in KKVP. The N-body results
are normalized to the volume within 200 $\hkpc$.
\label{fig:submf}}
\end{inlinefigure}

We see that the semi-analytic merger trees produce more of the small
sub-halos ($V_{\rm max} \lsim 20$ km/s) than the ART simulations, as
found by BKW00. The simulations of M99 have a factor of $\sim $2--3
more of the smallest halos than the ART simulations, but this is
likely to be mainly due to the $\om=1$ SCDM cosmology used for the M99
simulations. The recent \lcdm\ simulations presented by Font \&
Navarro (2001) and Governato et al. (2001) both appear to have about
40 \% less substructure than the SCDM simulations of M99, in good
agreement with the ART results. Some of the discrepancy could also be
due to differences in the algorithm used to identify sub-halos, or to
the slightly different masses of the host halos.

BKW00 adopted a model for tidal destruction of sub-halos in order to
bring the number of surviving sub-halos in the semi-analytic merger
trees into agreement with that found in the ART simulations. While it
is clear that tidal destruction of sub-halos does occur in the
simulations (A. Kravtsov, private communication), and tidal
destruction of dwarf galaxies does occur in the Local Group (we are
witnessing it in progress in the case of the Sagittarius dwarf) it is
still possible that dissipationless simulations overestimate the
effects of tidal destruction relative to the real Universe, where the
presence of baryons might make galaxies more robust to disruption.  We
opt to neglect the effects of tidal disruption in this work, for this
reason and also because we wish to find out if the effects of
photoionization squelching alone can solve the sub-structure problem.

\subsection{Photoionization Squelching}
The basic ingredients of the models used here, which include gas
cooling, star formation, supernovae feedback, chemical evolution and
stellar population synthesis, are described in detail in SP and
Somerville, Primack \& Faber (2001). We introduce one new ingredient,
the squelching of gas infall by a photoionizing background. We assume
that the Universe becomes reionized instantaneously at a redshift
$z_{\rm ion}$, which we treat as a free parameter. If a halo collapses
before $z_{\rm ion}$, it is allowed to capture a mass $M_g = f_b
M_{\rm vir}$ of gas which subsequently becomes available for cooling
and star formation, where $f_b$ is the universal baryon fraction and
$M_{\rm vir}$ is the virial mass of the halo. After reionization, the
gas captured by a halo is given by the fitting function provided by
Eqn.~7 of G00:
\begin{equation}
M_g = \frac{f_b M_{\rm vir}}{[1+0.26 M_C/M_{\rm vir}]^3}
\end{equation}
where $M_C$ is a characteristic mass, which represents the mass of
objects that on average retain 50\% of their gas. We approximate $M_C$
as the mass corresponding to a halo with virial velocity $V_C=50$ km/s
collapsing at redshift $z$ (see the appendix of SP). This provides a
good description of the redshift dependence of $M_C$ found in the
simulations of G00: $M_C$ increases from about $10^8 \msun$ at $z=8$
to several $10^{10} \msun$ at $z=0$. Then, in agreement with the
results shown in Fig.~2 of G00 and with previous results (e.g. Thoul
\& Weinberg 1996; Quinn et al. 1996), we find that halos with virial
velocity $V_c \sim 50$ km/s are able to capture about half of the gas
that they would accrete in the absence of a background, while halos
with $V_c<30$ km/s are able to accrete very little gas, and halos with
$V_c > 75$ km/s are nearly unaffected.

\begin{inlinefigure}
\begin{center}
\resizebox{\textwidth}{!}{\includegraphics{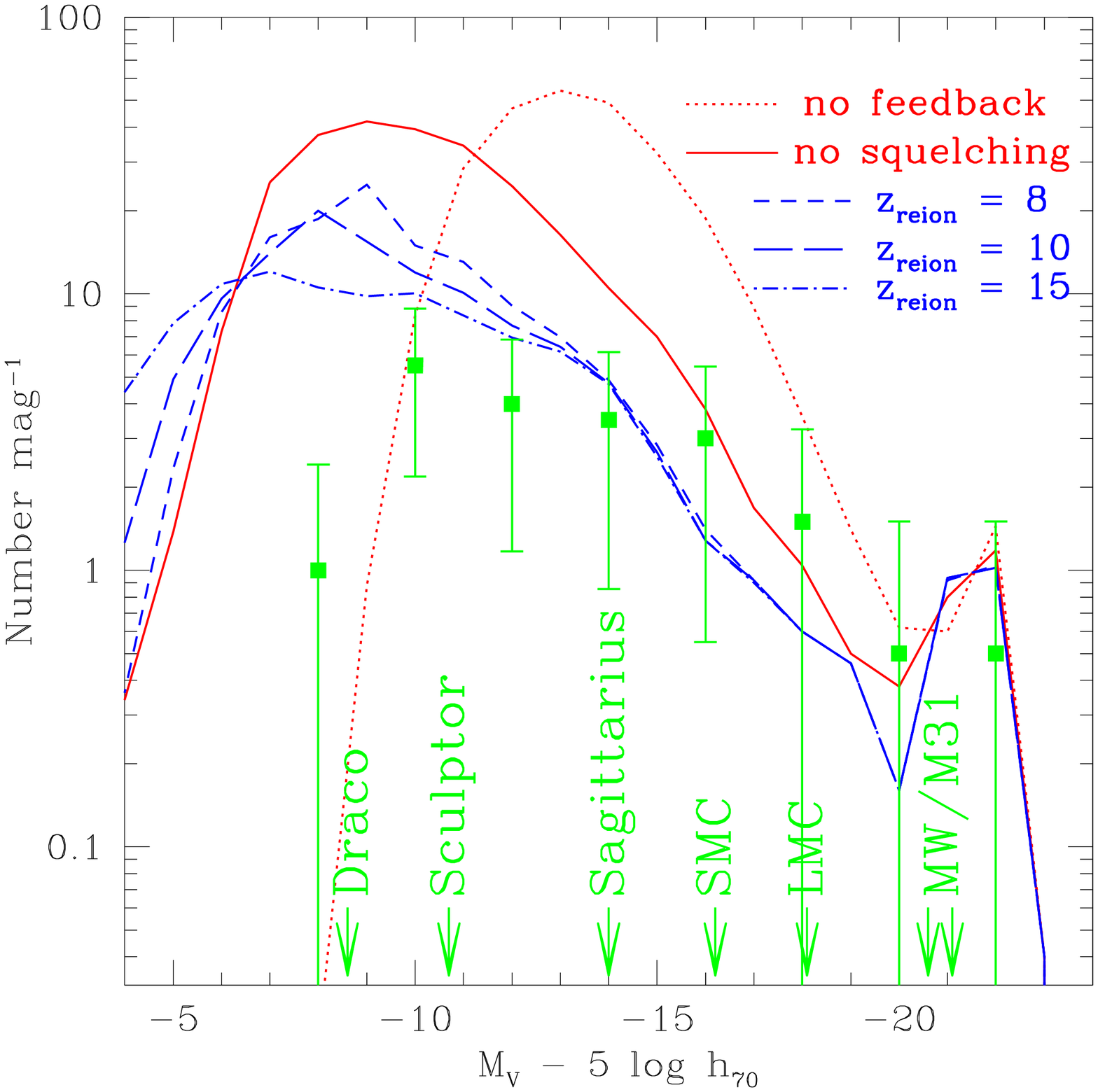}}
\end{center}
\figcaption{The luminosity function of the Local Group. Symbols with
error bars show the observed local group luminosity function, with
Poisson errors. The dotted line shows the results of the models with
no feedback of any kind, and the solid line shows the results of the
models with supernovae feedback but no photoionization
squelching. Short dashed, long dashed, and dot-dashed lines show the
model results with squelching, for three different assumed epochs of
reionization as indicated on the figure panel. The magnitudes of
several familiar Local Group members are indicated.
\label{fig:mwlf}}
\end{inlinefigure}

We normalize our models to reproduce the V-band luminosity of the
central Milky Way or M31-like galaxies in our ``Local Group'', which
we represent as two equal mass halos each with $M_{\rm vir} = 10^{12}
\msun$. The luminosity function of dwarf satellite galaxies for an
ensemble of such halos is shown in Fig.~\ref{fig:mwlf}. The data
points show the V-band luminosity function of dwarf galaxies in the
Local Group from the compilation of 
Irwin\footnote{http://www.ast.cam.ac.uk/\~{ }mike/local\_members.html} (see
also Mateo 1998).  We show the model results without supernovae (SN)
feedback or squelching, and for the models including supernovae
feedback according to the usual treatment (e.g. Kauffmann, White \&
Guiderdoni 1993) but without photoionization squelching.  For the
models without SN feedback, the fall-off at $M_V \lsim -13$ is due to
the fact that gas within these halos cannot cool efficiently via
atomic processes, and we have not included cooling by molecules, which
we expect to be inefficient (Machacek, Bryan \& Abel 2001; Haiman,
Abel \& Rees 2000). Including SN feedback pushes this ``bend'' to
fainter magnitudes.  The models including supernovae feedback only, at
a level similar to previously published semi-analytic models, fit the
luminosity function of satellite galaxies down to about $M_V \simeq
-15$ (a bit fainter than the SMC), but overproduce fainter objects by
a considerable factor --- more than one order of magnitude for the
faintest objects (such as Draco and Carina).

We now consider the effect of photoionization squelching as discussed
above. Results are shown for three different values of the
reionization epoch $z_{\rm ion}$, and are seen to be relatively
insensitive to this parameter within a reasonable range of values
(note that observations constrain the epoch of reionization to $6 \la
z_{\rm reion} \la 40$, and most CDM-based modelling suggests that it
lies in the range $6 \la z_{\rm reion} \la 20$; see the recent reviews
by Shapiro (2001) and Barkana \& Loeb (2001) and references therein).
The large galaxies (the ``Milky Way'' and ``M31'') are virtually
unaffected, the LMC-sized galaxies are slightly affected, and galaxies
smaller than the SMC are significantly affected by the presence of the
background.  The models now produce a reasonably good fit to the
observed dwarf satellite luminosity function, especially given that
tidal destruction may further reduce the number of satellites and that
the observed LF may be incomplete for the faintest objects.

\section{Discussion and Conclusions}

We have compared the multiplicity function of sub-halos within a
``Milky Way'' sized halo predicted by semi-analytic Monte Carlo merger
trees with that found in high-resolution N-body simulations. In
agreement with BKW00, we found that semi-analytic merger trees predict
more substructure than is found in the ART simulations. This is
probably due to the tidal destruction of sub-halos in the N-body
simulations, which we have chosen to neglect in the semi-analytic
models. In continuing to assess the sub-structure issue, it will be
important to determine the importance of tidal destruction of galaxies
when the presence of baryons and the effects of baryonic contraction
on the halo are included. The simulations of M99 have a factor of 2--3
more sub-halos than the ART simulations, which is probably due to the
SCDM cosmology used by M99. The physical origin of this cosmology
dependence (now confirmed by several groups) is not yet clear and
should be investigated.

All previously published studies of which we are aware (e.g. M99;
KKVP99; BKW00) have illustrated the sub-structure problem in terms of
the circular velocity function of sub-halos and satellite
galaxies. However, in order to do this the observed velocity
dispersions of the dwarf satellite galaxies (which are predominantly
spheroidal) must be converted to rotation velocities by making some
assumptions about the geometry and the potential. It is generally
assumed that the orbits are isotropic and the potentials isothermal
(M99); however, this assumption is uncertain and probably inaccurate
at the 50\% level, as the velocity dispersions of many spheroids drop
in the inner region where the dwarf velocity dispersions are measured
(White 2000). In addition, the presence of baryons modifies the
rotation curve, so that the peak velocity of the dark matter halo does
not necessarily correspond with the peak velocity of the galaxy that
forms within it. Moore (2001) and Font \& Navarro (2001) attempt to
account for these effects, and it appears unlikely that these
corrections are large enough to make the sub-structure problem go
away. Still, it is useful to present the comparison in a different
form, in terms of the predicted and observed luminosity function of
dwarf satellites. While many uncertain ingredients go into modeling
galaxy luminosities, at least this quantity is directly observed and
the selection effects are relatively well understood. We find, in
agreement with the previous studies based on velocity functions, that
models without any kind of feedback dramatically overpredict the
number of dwarf satellite galaxies in the Local Group.

We find that the standard implementation of supernovae feedback
generally used in semi-analytic galaxy formation models cannot, by
itself, solve the substructure problem. If we increase the efficiency
of SN feedback to the point where enough dwarf galaxies are suppressed
to solve the sub-structure problem, the feedback produces too great an
effect on larger galaxies, introducing a curvature in the Tully-Fisher
relation that is incompatible with observations (see SP; also Moore
2001). Supernovae feedback might be made to work by introducing a more
complicated mass dependence (rather than a power-law with a single
slope), but this would have to be formulated in a purely ad-hoc way,
and still might have difficulties with other observational
constraints. 

When we include the suppression of gas accretion by small halos after
reionization in our semi-analytic model of galaxy formation, using a
recipe based on state-of-the-art hydrodynamic simulations, we find
that with no additional tuning, the model then produces very good
agreement with the observed luminosity function of satellite galaxies
in the Local Group. The properties of luminous galaxies in the models
(which have been shown previously to agree well with observations) are
nearly unaffected. The results are not very sensitive to the redshift
of reionization, which we treat as a free parameter, although
squelching is slightly less effective (leading to survival of more
dwarf galaxies) when reionization is assumed to be later, because more
galaxies are able to form before the squelching is turned on.

It should also be noted that here we have considered only one possible
form of photoionization feedback, the suppression of gas collapse due
to a uniform background produced by distant stars or quasars. We have
neglected several further effects. Even gas that has managed to
collapse might be ``boiled'' out of very small halos (Barkana \& Loeb
1999; Haiman, Abel \& Madau 2001). This effect is expected to be
important mainly in the smallest of the halos harboring observable
dwarf galaxies ($V_c \lsim 20$ km/s). In addition, the presence of the
background would reduce the efficiency of atomic cooling.
Furthermore, the effect of turning on a quasar or starburst in close
proximity to the dwarf halos (as for example, if our Galaxy or M31
underwent such episodes, which presumably they did at some point in
the past, since both have bulges and probably black holes) might lead
to a more dramatic form of photoionization feedback than the one we
have considered here. Some combination of these effects might help to
reduce the remaining excess of very faint ($M_V \simeq -8$) satellites
in our models. 

For $M_V \lsim -10$, the models with squelching come very close to
producing the correct number of dwarf galaxies even neglecting the
above processes and \emph{with no tidal destruction}. If our treatment
of the efficiency of squelching, supernovae feedback, and star
formation is reasonably accurate, this suggests that tidal destruction
may not be as efficient as dissipationless simulations (e.g. BKW00)
would suggest --- otherwise, our models would not produce
\emph{enough} dwarf galaxies! Similarly, models with reduced power on
small scales (such as Warm Dark Matter) may actually suffer from a
reverse sub-structure problem, and fail to produce a sufficient number
of dwarf satellite galaxies.

The idea that we have explored here is very similar to that
investigated by BKW00. However, BKW00 did not explicitly model any gas
processes (cooling, star formation, SN feedback, etc) and so were
forced to introduce a more simplified criteria to determine which
sub-halos would harbor observable galaxies.  While our models use
simple recipes to represent the relevant physics, they do contain a
self-consistent treatment of the many interwoven processes that
determine the present day luminosity of a dwarf galaxy. We therefore
consider it encouraging that our investigation has produced similar
conclusions to those of BKW00.

While there have been suggestions that the photoionization phenomenon
could be responsible for producing Lyman-limit QSO absorption systems
at $z>3$ (Abel \& Mo 1998) and bursting dwarf galaxies at $z\sim1$
(Babul \& Rees 1992), and the effects of photoionization feedback on
the epoch of reionization itself have been explored (Haiman \& Loeb
1997; Ciardi et al. 2001), there have been relatively few
investigations connecting this physics with the properties of nearby
dwarf galaxies, perhaps because most semi-analytic galaxy formation
models have not included this effect to date. With this mechanism in
place, we are now in a position to investigate many other properties
of dwarf galaxies within this framework.

The squelching scenario has a number of interesting implications. It
suggests that there are many barren halos floating around in the Local
Group, as well as many very faint galaxies that lie below current
detection limits. There has been some concern that these halos might
heat the disk of our Galaxy beyond allowed observational limits, but
recent studies (Vel\'{a}zquez \& White 1999; Font \& Navarro 2001)
suggest that the halos in question are too small to do much
damage. However, the prospects of detecting these halos via
gravitational lensing or by direct detection if the Earth should
happen to pass through one of them (Moore et al. 2001) remain
intriguing possibilities.

%===================================
\section*{Acknowledgments}
\begin{small}
We thank B. Moore, A. Klypin, and A. Kravtsov for making the results
of their simulations available to us electronically, and for useful
discussions.  We also thank R. Barkana for useful discussions, and
T. Abel, J. Bullock, and A. Kravtsov for helpful comments on the
manuscript.  RSS is supported by a rolling grant from PPARC.
\end{small}
%=====================================

\def\re{\reference}


\begin{references}

\reference{abel-mo} Abel, T. \& Mo, H.J. 1998, ApJ, 494L, 151

\reference{babul} Babul, A., \& Rees, M.J. 1992, MNRAS, 255, 346

\reference{barkana} Barkana, R., Haiman, Z. \& Ostriker, J.P. 2001, 
astro-ph/0102304

\reference{barkana_loeb} Barkana, R., \& Loeb, A. 2001, Physics
Reports, in press, astro-ph/0010468

\reference{bode} Bode, P., Ostriker, J.P., \& Turok, N. 2001, ApJ,
556, 93

\reference{bullock} Bullock, J.S., Kolatt, T.S., Sigad, Y.,
Somerville, R.S., Kravtsov, A.V., Klypin, A., Primack, J.R., \& Dekel,
A. 2001, MNRAS, 321, 559

\reference{bkw} Bullock, J.S., Kravtsov, A.V, Weinberg, D.H., 2000, 
ApJ, 548, 33 (BKW00)

\reference{cole} Cole, S., Lacey, C.G., Baugh, C.M., \& Frenk,
C.S. 2000, MNRAS, 319, 168

\reference{chiu} Chiu, W., Gnedin, N.Y., \& Ostriker, J.P. 2001, 
astro-ph/0103359

\reference{ciardi} Ciardi, B. Ferrara, A., Governato, F., Jenkins,
A. 2000, MNRAS, 2000, 314, 11

\reference{efs} Efstathiou, G. 1992, MNRAS, 256, 43

\reference{fn} Font, A.S., \& Navarro, J.F. 2001, astro-ph/0106268

\reference{gnedin} Gnedin, N.Y. 2000, ApJ, 542, 535 (G00)

\reference{gnedin} Gnedin, O.Y. \& Ostriker, J.P. 2001, 
astro-ph/0010436

\reference{har} Haiman, Z., Abel, T., \& Rees, M.J. 2000, 534, 11 

\reference{ham} Haiman, Z., Abel, T., \& Madau, P. 2001, ApJ, 551, 599

\reference{hl} Haiman, Z., \& Loeb, A. 1997, ApJ, 483, 21

\reference{haiman} Haiman, Z., Thoul, A.A., \& Loeb, A. 1996, ApJ, 464, 523

\reference{hogan} Hogan, C.J., \& Dalcanton, J.J. 2000,
Phys. Rev. D62, 063511

\reference{hui} Hui, L. 2001, Phys. Rev. Lett. 86, 3467

\reference{ikeuchi} Ikeuchi, S. 1986, AP\&SS, 118, 509

\reference{kaplinghat} Kaplinghat, M., Knox, L., \& Turner, M.S. 2000, 
astro-ph/0005210

\reference{kam} Kamionkowski, M. \& Liddle, A.R. 2000,
Phys. Rev. Lett. 84, 4525

\reference{kwg} Kauffmann, G., White, S.D.M., \&
Guiderdoni, B., 1993, MNRAS, 264, 201

\reference{kkvp} Klypin, A.A., Kravtsov, A.V., Valenzuela, O., 
\& Prada, F. 1999, ApJ, 522, 82 (KKVP)

\reference{art} Kravtsov, A.V., Klypin, A.A., \& Khokhlov, A.M. 1997,
ApJS, 111, 73

\reference{machacek} Machacek, M.E., Bryan, G.L., \& Abel, T. 2001,
548, 509

\reference{mateo} Mateo, M.L. 1998, ARA\&A, 36, 435

\reference{miralda} Miralda-Escud\'{e}, J., 2000, astro-ph/0002050

\reference{moore:01b} Moore, B., Calcaneo-Roldan, C., Stadel, J.,
Quinn, T., Lake, G., Ghigna, S., Governato, F. 2001, astro-ph/0106271

\reference{moore:01} Moore, B., 2001, Plenary talk, 20th Texas
Symposium, eds. J. C. Wheeler \& H. Martel, AIP Conference Series,
astro-ph/0103100

\reference{moore:99} Moore, B., Ghigna, F., Governato, F., Lake, G., 
Stadel, J., \& Tozzi, P. 1999, ApJ, 524, L19 (M99)

\reference{ns} Navarro, J., \& Steinmetz, M. 1997, ApJ, 478, 13

\reference{ps} Press, W.H., \& Schechter, P. 1974, ApJ, 187, 425

\reference{quinn} Quinn, T., Katz, N., \& Efstathiou, G. 1996, ApJ, 278, 49

\reference{rees} Rees, M.J., 1986, MNRAS, 218, 25

\reference{shapiro} Shapiro, P. 2001, Plenary Session, 
20th Texas Symposium, eds. J.C. Wheeler \& H. Martel, AIP
Conference Series, astro-ph/0104315

\reference{shapiro} Shapiro, P.R., Giroux, M.L., \& Babul, A. 1994, 
ApJ, 427, 25

\reference{spergel} Spergel, D.N., \& Steinhardt, P.J. 2000,
Phys.Rev.Lett., 84, 3760

\reference{sk} Somerville, R.S. \& Kolatt, T.S. 1999, MNRAS, 305, 1

\reference{sp} Somerville, R.S. \& Primack, J.R. 1999, MNRAS, 310, 1087

\reference{spf} Somerville, R.S., Primack, J.R., \& Faber, S.M., 2001,
MNRAS, 320, 504

%\reference{tb} Taylor, J. \& Babul, A., 2001, astro-ph/0012305

\reference{tw} Thoul, A.A. \& Weinberg, D.H. 1996, ApJ, 465, 608

\reference{vw} Vel\'{a}zquez, H., \& White, S.D.M. 1999, MNRAS, 304, 254

\reference{weinberg} Weinberg, D.H., Hernquist, L., \& Katz, N. 1997, 
ApJ, 477, 8

\reference{white_itp} White, S.D.M. 2000, ITP conference, 
url:online.itp.ucsb.edu/online/galaxy\_c00

\end{references}
\end{document}